\documentclass[preprints,article,accept,oneauthor,pdftex]{Definitions/mdpi}

\firstpage{1}
\pubvolume{1}
\issuenum{1}
\articlenumber{0}
\pubyear{2026}
\copyrightyear{2026}
\datereceived{}
\dateaccepted{}
\datepublished{}
\hreflink{https://doi.org/} 

\Title{TEPCat: the Transiting Extrasolar Planet Catalogue}

\TitleCitation{TEPCat: the Transiting Extrasolar Planet Catalogue}

\Author{John Southworth\orcidA{}}

\AuthorNames{John Southworth}

\AuthorCitation{Southworth, J.}

\address{Astrophysics Group, Keele University, Staffordshire, ST5 5BG, UK; taylorsouthworth@gmail.com}

\corres{Correspondence: taylorsouthworth@gmail.com}

\abstract{Transiting extrasolar planets are extraordinarily valuable for understanding the characteristics and formation of planets, because they are the only exoplanets whose physical and orbital properties can be measured to high precision. Thousands are now known, and it is important to maintain a database of them for use by the scientific community. TEPCat performs this task: it is a critical compilation of the physical and observable properties of the known transiting planetary systems. This work introduces the motivation for TEPCat, its scope, contents, and implementation. Example plots of interesting quantities are constructed. The classification of planets and of the eclipse features in their light curves is discussed. TEPCat can be found at \texttt{https://www.astro.keele.ac.uk/jkt/tepcat/}.}


\newcommand{\Teff}{\ensuremath{T_{\rm eff}}}                      
\newcommand{\Msun}{\ensuremath{\,{\rm M}_\odot}}                  
\newcommand{\Rsun}{\ensuremath{\,{\rm R}_\odot}}                  
\newcommand{\Mjup}{\ensuremath{\,{\rm M}_{\rm Jup}}}              
\newcommand{\Rjup}{\ensuremath{\,{\rm R}_{\rm Jup}}}              
\newcommand{\Mearth}{\ensuremath{\,{\rm M}_\oplus}}               
\newcommand{\Porb}{\ensuremath{P_{\rm orb}}}                      
\newcommand{\reff}[1]{{#1}}

\begin{document} 

\section{Introduction}

TEPCat (the Transiting Extrasolar Planet Catalogue) is an online catalogue of transiting extrasolar planets (TEPs) and their physical properties. It is regularly updated and can be found at \texttt{https://www.astro.keele.ac.uk/jkt/tepcat/}. It collects the observable properties of the systems (including sky position, apparent magnitudes and orbital ephemerides), the physical properties of the stars (including mass, radius and effective temperature) and planets (including mass, radius and equilibrium temperature), and orbital properties (such as period, semimajor axis, eccentricity and obliquity). All information is collated and made freely available in both webpage and machine-readable form, alongside relevant totals and plots. 

TEPCat is widely used by the community. Examples of this include studies on the orbital obliquity of planetary systems \cite{Attia+23aa,Rice+24aj,Rusznak+25apj,Rossi+25xxx}, provision of the physical properties of giant planets for the calculation of transmission spectra \cite{Goyal+18mn}, inclusion of physical properties to augment atmospheric parameters determined for a large sample of planet host stars \cite{Santos+13aa}, sample definition for transit timing work \cite{IvshinaWinn22apj}, orbital decay studies \cite{CameronJardine18mn}, and sample selection for population studies of transiting planets \cite{HatzesRauer15apj,Sarkis+21aa,Wang++26apj}. It has also been used to select targets for observation with the space missions Ariel \cite{Edwards+19aj}, Twinkle \cite{Edwards+19exa}, CHEOPS \cite{Borsato+21mn} and PLATO \cite{Eschen+24mn,Nascimbeni+25aa}, plus the definition of many other observational and theoretical projects (including the author's).

TEPCat was briefly described as part of another project \cite{Me11mn}. In this work we present it in detail, including motivation (Section\,\ref{sec:motivation}), its contents (Section \ref{sec:contents}) and scope (Section \ref{sec:scope}), implementation (Section\,\ref{sec:implementation}), and visualisation of some of the more interesting data included in the catalogue (Section\,\ref{sec:plots}). Sections \ref{sec:class} and \ref{sec:nomen} discuss the classification of planets and eclipse features, and we conclude in Section\,\ref{sec:conc}.


\section{Motivation}
\label{sec:motivation}

The growth of research about transiting extrasolar planets (TEPs) over the past 25 years is remarkable, but has been more than matched by the increase in the number of known planets. The first suggestion to search for these objects was made by Struve in 1952 \cite{Struve52obs}, in a prophetic paper only two pages long. Just one paragraph discussed transits; the majority of the paper was dedicated to showing that short-period massive planets might be detected by radial velocity (RV) with the photographic spectroscopy techniques then in use. In 1971 Rosenblatt \cite{Rosenblatt71icar} noted that planet transits could be identified by slightly colour changes of a star during transit due to the variation of the strength of limb darkening with wavelength. Borucki \& Summers \cite{BoruckiSummers84icar} considered the subject in more detail in 1984, and advocated for large surveys of bright stars. More effort was expended by the community on RV searches \cite{CampbellWalker79pasp,Campbell++88apj,Latham+89nat,DuquennoyMayor91aa,MarcyButler92pasp}, leading to the first detections of planets orbiting normal \reff{(main-sequence)} stars \cite{MayorQueloz95nat,MarcyButler96apj}. 

The first \emph{transiting} planet, HD\,209458\,b, was identified in 1999 by monitoring its host star around the time a transit might occur \cite{Henry+00apj,Charbonneau+00apj}; the planet had already been detected from RV measurements. It confirmed the supposition of Struve \cite{Struve52obs} that short-period giant planets exist and can produce detectable transit signatures. It took another three years for the second TEP to be discovered -- OGLE-TR-56 \cite{Konacki+03nat} from survey observations obtained by the OGLE project \cite{Udalski+02aca} -- and this was the first to be found by its transits rather than RV signature.

The idea of running photometric surveys to simultaneously search for transits around thousands of stars was originally put forward by Borucki \& Summers \cite{BoruckiSummers84icar}. The first materialisation of this concept was the Vulcan photometer run by Borucki et al.\ \cite{Borucki+01pasp} in the late 1990s. It observed 6000 bright stars in one field for three months, finding over 100 variable stars and demonstrating sufficient photometric precision to detect planet transits. Vulcan identified 35 transit candidates \cite{Latham03aspc} but -- unluckily -- none turned out to be true planets. Many more ground-based surveys began operating in the 2000s using similar survey designs \cite{Pepper++03aca}, and contributed to a major increase in the number of known planets; the most successful of the surveys were SuperWASP \cite{Pollacco+06pasp}, HATNet \cite{Bakos+02pasp} and HATSouth \cite{Bakos+13pasp} (see below).

The number of known TEPs increased hugely once dedicated survey telescopes were placed in space (see also ref.\ \cite{Me21univ}), the first being CoRoT \cite{Baglin+06conf}. This was followed by the \textit{Kepler} space telescope \cite{Borucki16rpph} and its extension as the K2 mission \cite{Howell+14pasp}, which was used to discover thousands of examples in relatively small fields of view (115\,deg$^2$). The faintness of the majority of the TEP candidates \reff{(median magnitude 14.5)} means confirmation of them has been difficult and often beyond current technological capabilities. The TESS mission \cite{Ricker+15jatis} solved this problem by observing brighter stars over most of the sky \reff{(the median magnitude of the TESS planet candidates is 11.8)}, with the drawback that only a small fraction of objects were observed for more than one month at a time. The forthcoming PLATO mission \cite{Rauer+25exa} will solve both problems by observing bright stars \reff{(almost all target stars are brighter than magnitude 13)} continuously for two years or more.

In 2007, when the author wrote a paper outlining a new method to determine the surface gravities of TEPs \cite{Me++07mn}, only 14 were known. It was nevertheless clear that this number would greatly increase, so we began a series of papers presenting homogeneous reanalyses of TEPs based on existing and new data from both ground-based and space-based telescopes. These \textit{Homogeneous Studies} series of papers \cite{Me08mn,Me09mn,Me10mn,Me11mn,Me12mn} eventually included 82 TEPs, all analysed using methods developed to carefully account for both random and systematic errors in both the data and the theoretical models needed to measure the physical properties of the system. Further systems were added from a series of papers presenting high-quality photometry from telescope defocussing \cite{Me+09mn,Me+09mn2,Me+09apj,Me+10mn,Me+11aa,Me+12mn,Me+12mn2,Me+12mn3,Me+13mn,Me+14mn,Me+15mn,Me+15mn2,Me+16mn,Me+18mn} as well as those from collaborators using the same methods \cite{TregloanMe13mn,Ciceri+13aa,Mancini+13mn,Mancini+13mn2,Mancini+13aa,Mancini+14mn,Mancini+14aa,Mancini+14aa2,Ciceri+15aa2,Mancini+15aa,Tregloan+15mn,Mancini+16aa,Evans++16apj,Ciceri+16mn,Mancini+16mn,Mancini+16mn2,Mancini+17mn,Mancini+19mn,Mancini+22aa}. In 2010 it was clear that it would be useful to compile these homogeneously-determined properties and make them freely available, rather than expecting others to do this for themselves. The author therefore developed a website to describe and present these properties. It also included a catalogue of properties for all other known TEPs taken from the published literature, and therefore inhomogeneous, which had been compiled to aid work on the \textit{Homogeneous Studies}. This is TEPCat.

Two major changes have been made to TEPCat since then. The first was to split the catalogue into ``well-studied'' and ``little-studied'' objects, motivated by the publication in one paper of 851 new planets \cite{Rowe+14apj}. These were characterised using \textit{Kepler} data and a small amount of ground-based observations (high-resolution imaging to rule out blends, and spectra to measure the properties of the host star), so were measured to relatively low precision and had unknown masses. Simply adding these to TEPCat would have swamped the main webpages and made it harder to find information on TEPs that had been studied in detail. The ``little-studied'' category was therefore created to include these data in TEPCat but in separate tables. \reff{The second change was the removal of the \textit{Homogeneous Studies} results in 2023, as they were no longer useful for two reasons. First, most were based on data which have now been superseded in quality, primarily by the TESS mission. Second, newer analysis methods are now available, such as more modern theoretical stellar models and direct measurement of stellar radius using data from the \textit{Gaia} satellite.}


\section{Data included in TEPCat}
\label{sec:contents}

TEPCat is designed to hold sufficient information to support studies concerned with the properties of TEPs or target selection for new observing projects. This information is separated into multiple tables for ease of display, but a downloadable file of all information for each TEP is also available to avoid people having to join tables together.

The \texttt{Observables} table contains all properties which are directly useful for performing an observational project: sky position, apparent magnitude in the $V$ and $K$ passbands, orbital ephemeris, and the transit duration and depth. It also includes the date on which the planet's nature was confirmed, to allow for cumulative plots (see later), and the source of the data in which the transits were discovered (usually the name of an observing project or space telescope). Table\,\ref{tab:obs} lists the information in the \texttt{Observables} table. The reference for the ephemeris is given using the 19-character bibcode convention adopted by NASA ADS (\texttt{https://ui.adsabs.harvard.edu/}), and also linked in the \texttt{html} file to the ADS page for the reference.

The orbital ephemeris in the \texttt{Observables} table must be linear, i.e.\ correspond to a constant orbital period. There are multiple phenomena which can modify the orbital period and thus change the times of transit, including orbital decay (e.g.\ WASP-12 \cite{Hebb+09apj,Maciejewski+13aa}), gravitational interactions \cite{HolmanMurray05sci}, and the light-time effect due to additional stars or planets in the system (e.g.\ HAT-P-26 \cite{Hartman+11apj2,Vonessen+19aa}). Inclusion of such effects would require much effort and would only apply to a small minority of objects, so in all cases the best linear ephemeris is reported instead. Anyone using TEPCat to plan observations for which precise timings are necessary should check the literature in advance for reports of transit timing variations.

\begin{specialtable}[t] 
\caption{Information given in the \texttt{Observables} table in TEPCat. The ``$\sigma$'' column indicates if an uncertainty 
is given for the quantity: in all cases these are a single value so do not account for asymmetric error bars. \label{tab:obs}}
\renewcommand{\arraystretch}{1.3}
\begin{tabular}{p{15mm}p{25mm}p{10mm}p{60mm}}
\toprule
\textbf{Symbol} & \textbf{Unit} & \textbf{$\sigma$} & \textbf{Description} \\
\midrule   
--              & --            & -- & Name of planet (including `b' or `c' only when necessary) \\
--              & --            & -- & Type: TEP for planet, BD for transiting brown dwarf, CBP for circumbinary planet) \\
--              & YYYY MM DD    & -- & Date planetary nature was confirmed (Gregorian calendar) \\
--              & --            & -- & Source of data from which planetary nature was deduced \\
$\alpha$        & h m s         & -- & Right ascension (normally J2000) \\
$\delta$        & d m s         & -- & Declination (normally J2000) \\
$V$             & mag           & -- & Apparent magnitude of the system in the $V$-band \\
$K$             & mag           & -- & Apparent magnitude of the system in the $K$- or $K_s$-band \\
$\tau_{4-1}$    & d             & -- & Transit duration \\
$\Delta F/F$    & \%            & -- & Transit depth \\
$T_0$           & BJD or HJD    & y  & Time of midpoint of reference transit (using the same time system as given in the original source) \\
\Porb           & d             & y  & Orbital period \\
--              & --            & -- & Reference for ephemeris \\
\bottomrule
\end{tabular}
\end{specialtable}

The \texttt{Properties} table gives all the physical properties of the TEPs and their host stars. All measured quantities are specified with separate upper and lower error bars. If any property (such as planet mass) has only an upper limit, this is indicated by setting the value of the quantity to zero and the upper error bars to the published upper limit. Upper limits are taken from the literature so cannot be made consistent: the most commonly quoted limits are 1$\sigma$, 3$\sigma$ and 95\%. Table\,\ref{tab:prop} lists the information in the \texttt{Properties} table. References are specified as indicated above. Two references are given: that for the discovery paper and that for the most recent paper from which one of the physical properties were taken.

There are several cases where a TEP system has been independently discovered by several consortia. If so, the first paper to appear as a journal article or on the arXiv preprint server is given precedence. If two (or more) of the discovery papers appear on the same day (as memorably happened for HD\,80606 \cite{Naef+01aa,Moutou+09aa,GarciaMccullough09apj,Fossey++09mn}), precedence is given to the paper with the best data and/or results; personal judgement is exercised in making this choice.

Some measurements are not published but are easily calculable from those which are. TEPCat contains the calculated values in the webpages, italicised and enclosed in brackets for easy identification, but not in the datafiles. The calculations implemented in TEPCat are:
\begin{itemize}
\item $a$ from $M_{\rm A}$, $M_{\rm b}$ (optional) and \Porb;
\item $\log g_{\rm A}$ from $M_{\rm A}$ and $R_{\rm A}$;
\item $\rho_{\rm A}$ from $M_{\rm A}$ and $R_{\rm A}$;
\item $g_{\rm b}$ from $M_{\rm b}$ and $R_{\rm b}$;
\item $\rho_{\rm b}$ from $M_{\rm b}$ and $R_{\rm b}$;
\item $T_{\rm eq}$ from \Teff, $R_{\rm A}$ and $a$.
\end{itemize}

\begin{specialtable}[t] 
\caption{Information given in the \texttt{Properties} table in TEPCat. The ``$\sigma$'' column indicates if 
uncertainties are given for the quantity: in all cases these are separate upper and lower error bars. \label{tab:prop}}
\renewcommand{\arraystretch}{1.3}
\begin{tabular}{p{15mm}p{15mm}p{10mm}p{70mm}}
\toprule
\textbf{Symbol}   & \textbf{Unit}   & \textbf{$\sigma$} & \textbf{Description} \\
\midrule   
--                & --              & --  & Name of planet (exactly matches the name in the \texttt{Observables} table) \\
\Teff             & K               & y   & Effective temperature of host star \\
\verb+[M/H]+      & dex             & y   & Metallicity of the host star \\
\Porb             & d               & --  & Orbital period (truncated to 3 d.p.) \\
$a$               & au              & y   & Semimajor axis of the relative orbit \\
$e$               &                 & y   & Orbital eccentricity \\
$M_{\rm A}$       & M$_\odot$       & y   & Mass of host star \\
$R_{\rm A}$       & R$_\odot$       & y   & Radius of host star \\
$\log g_{\rm A}$  & cgs             & y   & Logarithmic surface gravity of host star \\
$\rho_{\rm A}$    & $\rho_\odot$    & y   & Mean density of host star \\
$M_{\rm b}$       & M$_{\rm Jup}$   & y   & Mass of planet \\
$R_{\rm b}$       & R$_{\rm Jup}$   & y   & Radius of planet \\
$g_{\rm b}$       & m s$^{-2}$      & y   & Surface gravity of planet \\
$\rho_{\rm b}$    & $\rho_{\rm Jup}$& y   & Mean density of planet \\
$T_{\rm eq}$      & K               & y   & Equilibrium temperature of planet \\
--                & --              & --  & Reference for discovery paper \\
--                & --              & --  & Reference for physical properties \\
\bottomrule
\end{tabular}
\end{specialtable}

It is often the case that quantities are published in different units to those used in TEPCat: the most common are $M_{\rm b}$ in M$_\oplus$ instead of M$_{\rm Jup}$, $R_{\rm b}$ in R$_\oplus$ instead of R$_{\rm Jup}$, $g_{\rm b}$ given logarithmically, and $\rho_{\rm A}$ and $\rho_{\rm b}$ in g\,cm$^{-3}$ instead of $\rho_\odot$ or $\rho_{\rm Jup}$. In such cases, the values (and uncertainties) are converted to the correct units for inclusion in TEPCat. The physical constants used are those recommended in Resolutions 2012 B2 and 2015 B3 passed by the International Astronomical Union (IAU) \cite{Prsa+16aj}. Conversion factors between properties in the Solar system are typically known to much higher precision than needed for studying exoplanets. 

The unit R$_{\rm Jup}$ used in TEPCat requires further discussion. It is standard practise in the exoplanet community (and in TEPCat) to define this to be the equatorial radius of Jupiter, which is 71\,492\,km. However, a small fraction of publications instead report planetary sizes in units of the volume-mean radius of Jupiter, which is 69\,911\,km. The two values are different because Jupiter is a fast rotator (its rotation period is approximately 10 hours) and thus oblate. These values of the radii of Jupiter come from radio-occultation data obtained in the 1970s by the Voyager and Pioneer missions \cite{Lindal+81jgr} and have uncertainties of $\pm$4\,km. A new measurement became available just before the current manuscript was resubmitted for publication: Galanti et al.\ \cite{Galanti+26natas} measured values of $71\,488 \pm 0.4$\,km (eqatorial) and $69\,886 \pm 0.4$\,km (mean). The difference amounts to 0.0056\%, which is negligible for current measurements. Nevertheless, we recommend that the community continues to use the canonical value of 71\,492\,km for consistency, analogously to the physical properties of the Sun discussed in the previous paragraph.

The third and final table is the \texttt{Obliquities} table, which contains all known measurements of the sky-projected and true orbital obliquities of TEP systems. Whilst these are often taken to be properties of the planet, they are actually intrinsic properties of the system as they are angles between two different angular momentum vectors (those of the star's rotation and of the planet's orbit). Table\,\ref{tab:obl} lists the information in the \texttt{Obliquities} table. The references are given using author names instead of ADS bibcodes, but this anomaly is on list of things to change in TEPCat in the near future. Two different conventions exist for expressing obliquities: $\lambda$ and $\beta$ \cite{Ohta++05apj,Gimenez06aa,Triaud18book}. These differ only in direction, so $\lambda = -\beta$. The $\lambda$ convention predominates and is adopted in TEPCat: obliquities expressed as $\beta$ are converted to $\lambda$ for inclusion in the catalogue.

\begin{specialtable}[t] 
\caption{Information given in the \texttt{Obliquities} table in TEPCat. The ``$\sigma$'' column indicates if 
uncertainties are given for the quantity: in all cases these are separate upper and lower error bars. \label{tab:obl}}
\renewcommand{\arraystretch}{1.3}
\begin{tabular}{p{15mm}p{15mm}p{10mm}p{70mm}}
\toprule
\textbf{Symbol}   & \textbf{Unit}   & \textbf{$\sigma$} & \textbf{Description} \\
\midrule   
--                & --              & --  & Name of planet (exactly matches the name in the \texttt{Observables} table) \\
\Teff             & K               & y   & Effective temperature of host star \\
$\lambda$         & $^\circ$        & y   & Sky-projected orbital obliquity \\
$\psi$            & $^\circ$        & y   & True orbital obliquity \\
--                & --              & --  & Reference for measurement(s) \\
\bottomrule
\end{tabular}
\end{specialtable}

For the convenience of users, an extra table is available in TEPCat which collects all relevant information for each planet from the three tables described above. Each row contains 57 quantities for one planet and its host star. Like the other machine-readable datafiles, it is available in both \texttt{ascii} and \texttt{csv} formats.


\section{The scope of TEPCat}
\label{sec:scope}

TEPCat was originally designed to include the properties of all known TEPs. These properties have also been collected for transiting brown dwarfs (BDs) as well, because such systems are rare and interesting \cite{GretherLineweaver06apj,Vowell+25aj}. The two classes are labelled `TEP' and `BD' in the \texttt{Observables} table for ease of identification (the acronym `TBD' would have been better if it were not more widely used to indicate \textit{to be determined}). 

TEPs and BDs are divided by a mass limit of 13\Mjup, which is widely adopted in the scientific community. This represents the minimum mass required for thermonuclear fusion of deuterium in a celestial object's core, and must be calculated theoretically. Spiegel et al.\ \cite{Spiegel++11apj} found the limit to vary from approximately 11.0\Mjup\ to 16.3\Mjup\ depending on the chemical composition of an object and how much deuterium burning is deemed sufficient (e.g.\ 1\%, 50\%, 99\% or some other value). This complexity should be remembered by the reader, but complication is an enemy of catalogue maintenance so TEPCat adopts the simple approach of dividing the objects at 13\Mjup\ (ignoring the uncertainties in the $M_{\rm b}$ values). Further discussion of the definition of planets can be found in refs.\ \cite{Soter06aj,Chabrier+09conf,LecavelierLissauer22newar,Margot++24psj}.

An analogous situation exists for the division between BDs and stars, and can be conceptually separated by requiring sufficient mass for core hydrogen burning. Auddy et al.\ \cite{Auddy++16adast} found this limit to be between 67\Mjup\ and 91\Mjup, depending on chemical composition (again) and choices made when constructing theoretical models. This complexity is again important but unwelcome, so a dividing line is drawn at a mass of 75\Mjup: less massive objects are taken to be BDs or TEPs, and more massive objects are assumed to be stars. Thus transiting systems with $M_{\rm b} > 75$\Mjup\ (0.072\Msun) are considered to be eclipsing binary systems and are not included in TEPCat (but may be included in the author's catalogue of detached eclipsing binaies -- DEBCat -- instead \cite{Me15debcat}).

It is common for TEPs to have a known radius but not mass: this is a frequent occurrence in any photometric search for TEPs and can persist for low-mass planets whose orbital motion is too small to be measured spectroscopically using current technology. Whilst transiting objects with radii similar to Earth's cannot be anything but planetary, objects with radii similar to Jupiter could be planets, BDs, or even low-mass stars. A decision must therefore be made on whether to include a system in TEPCat if no constraint on $M_{\rm b}$ exists. The decision was made to include any system with $R_{\rm b} < 0.5$\Rjup, because they are unambiguously planetary, but not to include larger objects because their nature might be unclear.

There are also procedural criteria to be met for inclusion in TEPCat. It was  decided not to include any objects lacking sufficient information to be reobserved. All objects in TEPCat must therefore have a precise sky position (RA and Dec) and orbital ephemeris ($T_0$ and \Porb). Validated planets must have their planetary nature established to the 99.7\% (3$\sigma$) confidence level. Large compilations of new planets whose characterisation is limited and achieved with automated methods are a low priority for inclusion, due to the effort involved and their lesser scientific interest.


\section{Implementation of TEPCat}
\label{sec:implementation}

TEPCat was implemented in a hurry in 2011 to accompany the fourth \textit{Homogeneous studies} paper \cite{Me11mn}, so a quick and simple approach was taken. All underlying data are kept in a set of \texttt{ascii} files with fixed column widths, which has the benefits of visual clarity and speed of adding new results. These files are read into a monolithic IDL (Integrated Data Language, \texttt{https://www.nv5geospatialsoftware.com/Products/IDL}) \newline script for further processing. The script performs the following steps:
\begin{enumerate}
\item it resolves ADS bibcodes and arXiv identifiers into full URLs for the \texttt{html} pages;
\item it checks the input data for numbers outside the range of expected values;
\item it calculates a subset of parameters if they are not presented in the input files (see Section\,\ref{sec:contents});
\item it creates a set of machine-readable datafiles in \texttt{ascii} and \texttt{csv} format;
\item it creates a set of webpages containing all the data;
\item it creates a webpage with the total numbers of TEPs and BDs found from the various projects and missions;
\item it creates a webpage for each planet with all information available for it, some given in multiple units for convenience;
\item it creates a set of plots of interesting parameters (see below);
\item it makes a set of static webpages providing descriptions of TEPCat and further information.
\end{enumerate}

New results are identified via literature search and added when time permits. The arXiv preprint server (\texttt{https://arxiv.org/}) is manually checked every working day for relevant publications. Email lists from the major astrophysical journals (A\&A, AJ, ApJ, MNRAS, PASP) are obtained via subscription and checked on arrival for papers not submitted to arXiv, and for final ADS bibcodes of arXiv papers. The journal Nature is checked weekly. Other journals are not systematically checked due to the time and effort required. Updates and corrections are regularly received via email to the author.

\begin{figure}[t]
\includegraphics[width=13.5cm]{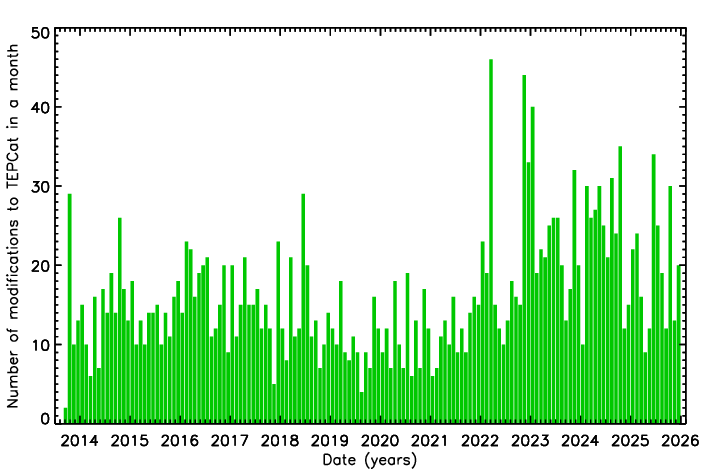}
\caption{Bar chart showing the number of modifications to TEPCat per month since late 2013, compiled from the \texttt{News} pages.
\label{fig:news}} \end{figure}

Older versions of TEPCat are systematically retained for scientific reproducibility. A snapshot of the catalogue is stored at the start of each month and made available in a directory with the naming convention YYYYmmm, e.g.\ ``2023mar'' for 1st March 2023. All changes since October 2013 are also itemised in the \texttt{News} section, which is split into individual years for faster loading. This last set of pages makes it possible to plot the number of changes as a function of time: Fig.\,\ref{fig:news} shows this in the form of a bar chart. A single modification is very varied and could be as small as the change or addition of one quantity (e.g.\ an orbital period or obliquity), or as big as the inclusion of 30 or more new planetary systems.


\section{Interesting results and plots}
\label{sec:plots}

Now the technical details have been outlined, we can discuss more interesting things. One of the more obvious uses of a database such as TEPCat is to plot various quantities of interest. This has been done below with the version of the database from 2026/02/01.

\begin{figure}[H]
\includegraphics[width=13.5cm]{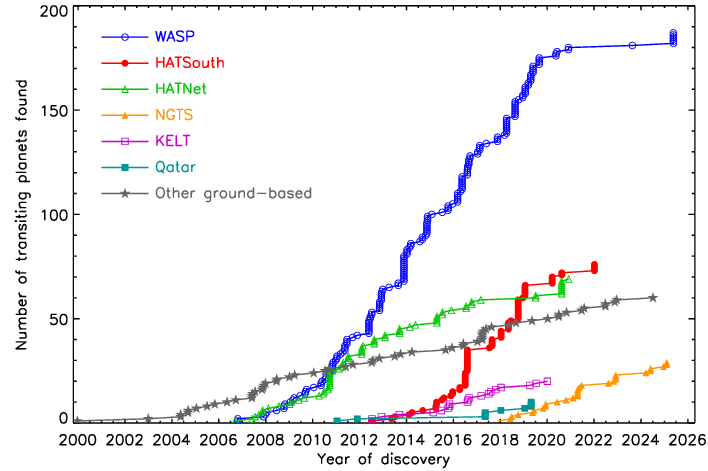} \\[10pt]
\includegraphics[width=13.5cm]{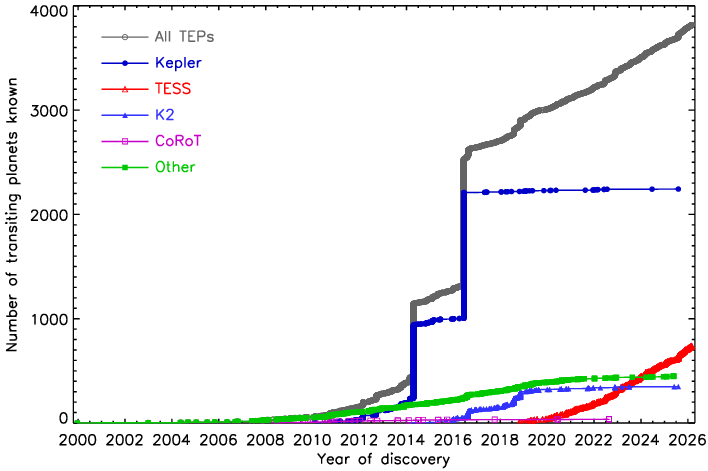} \\
\caption{Plots of the number of known TEPs as a function of time. The upper panel shows discoveries made using 
data from ground-based telescopes. The lower panel shows all discoveries, plus subdivisions into those from the 
\textit{Kepler}, TESS, K2 and CoRoT space missions. The keys give the symbols and colours used for each line. 
\label{fig:disc}} \end{figure}

Fig.\,\ref{fig:disc} shows the number of known TEPs as a function of time from the first discovery in late 1999 to the end of the year 2025. The date of discovery of each planet is taken to be the date when the article confirming its planetary nature became widely available through either arXiv or email alerts from the publishing journal. The history of discovery divides naturally into three phases, the last of which is ongoing. These phases are: \textit{(i)} the early discovery of small numbers of planets using a variety of telescopes; \textit{(ii)} increased discovery rates from large ground-based photometric surveys using small telescopes; and finally \textit{(iii)} large-scale discoveries using huge datasets from dedicated space telescopes. 

The ground-based discoveries (Fig.\,\ref{fig:disc}, top panel) are subdivided into individual consortia, restricted to those which announced the discovery of at least ten planets. The SuperWASP consortium dominates these discoveries, followed by the HATNet and HATSouth consortia. The lower panel of Fig.\,\ref{fig:disc} shows the total number of TEPs and separates out the four main space missions. The total number of TEPs is dominated by systems found in data from the \textit{Kepler} mission, which in turn come primarily from two publications validating 851 and 1284 planets \cite{Rowe+14apj,Morton+16apj}. The community-led discovery paradigm pioneered with the TESS mission produces a much smoother line in this plot.

\begin{figure}[H]
\includegraphics[width=13.5cm]{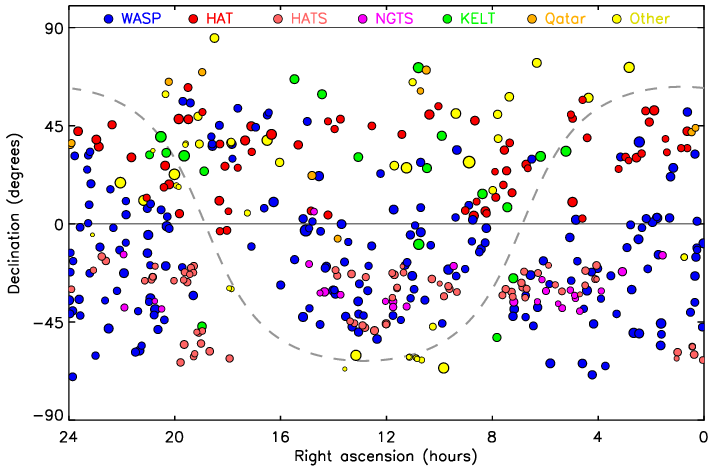} \\[10pt]
\includegraphics[width=13.5cm]{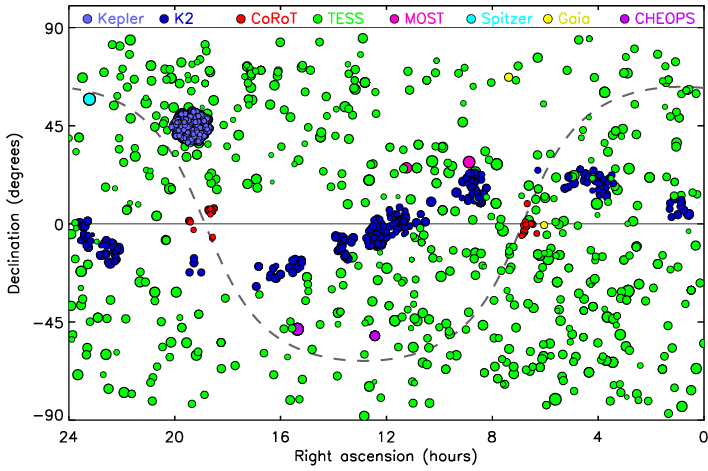} \\
\caption{Sky positions of the known transiting systems, in the equatorial coordinate system. The upper and lower panels 
show TEPs discovered using ground-based and space-based data, respectively. \reff{The point sizes are larger for brighter 
stars, with $V$-band apparent magnitudes ranging from 5.57 to 18.97.} The colours show which survey dataset led to the 
discovery, as indicated in the keys across the tops of the panels. The grey dashed line indicates the galactic plane.
\label{fig:skypos}} \end{figure}

Fig.\,\ref{fig:skypos} shows the sky positions of the objects in TEPCat, broken down into discoveries from ground-based (upper panel) and space-based (lower panel) data. The ground-based data are roughly isotropic, as expected for surveys of nearby bright stars. One feature that stands out is that there are few known near the galactic plane, because the greater stellar densities in these directions lead surveys to avoid this region. In contrast, many of the discoveries from space-based telescopes are concentrated in small regions of the sky due to survey design. All \textit{Kepler} discoveries were found in the one field observed during this mission. The K2 mission \cite{Howell+14pasp} found planets only near the ecliptic plane, because the \textit{Kepler} telescope was restricted to these directions by the failure of two of its four reaction wheels. Conversely, the CoRoT mission observed only towards the galactic centre and anti-centre to obtain continuous light curves for long time intervals. The TESS mission covers almost all the sky so its discoveries dominate the lower panel in Fig.\,\ref{fig:skypos}.

\clearpage

\begin{specialtable}[t] \centering
\caption{Numbers of transiting planets and brown dwarfs known. The numbers have been 
subdivided according to the data source used for the discovery of transits. The numbers
represent the contents of TEPCat on 2026/02/01. \label{tab:disc}}
\begin{tabular}{lrrr}
\toprule
\textbf{Data source}~~~~~~~~~~~~~~ & \textbf{TEPs} & \textbf{Transiting BDs} & \textbf{Reference}       \\
\midrule   
\textit{Kepler}      & 2237          & 5                       & \cite{Borucki16rpph}     \\
TESS                 & 708           & 29                      & \cite{Ricker+15jatis}    \\
K2                   & 345           & 4                       & \cite{Howell+14pasp}     \\
WASP                 & 185           & 2                       & \cite{Pollacco+06pasp}   \\
HATSouth             & 76            &                         & \cite{Bakos+13pasp}      \\
HATNet               & 69            &                         & \cite{Bakos+02pasp}      \\
CoRoT                & 33            & 4                       & \cite{Baglin+06conf}     \\
NGTS                 & 27            & 2                       & \cite{Wheatley+18mn}     \\
KELT                 & 19            & 1                       & \cite{Pepper+07pasp}     \\
Qatar                & 10            &                         & \cite{Alsubai+13aca}     \\
OGLE                 & 8             &                         & \cite{Udalski+03aca2}    \\
(other)              & 7             &                         & --                       \\
XO                   & 7             &                         & \cite{Mccullough+05pasp} \\
Spitzer              & 6             &                         & \cite{Werner+04apjs}     \\
MEarth               & 4             & 2                       & \cite{Irwin+09conf}      \\
TrES                 & 5             &                         & \cite{Alonso+04apj}      \\
TRAPPIST             & 3             &                         & \cite{Gillon+11conf}     \\
CHEOPS               & 2             &                         & \cite{Benz+21exa}        \\
\textit{Gaia}        & 2             &                         & \cite{Gaia16aa}          \\
MASCARA              & 2             &                         & \cite{Talens+17aa}       \\
MOST                 & 2             &                         & \cite{Walker+03pasp}     \\
Wendelstein          & 2             &                         & \cite{Obermeier+16aa}    \\
WTS                  & 2             &                         & \cite{Cappetta+12mn}     \\
GPX                  &               & 1                       & \cite{Benni+21mn}        \\
KPS                  & 1             &                         & \cite{Burdanov+16mn}     \\
POTS                 & 1             &                         & \cite{Koppenhoefer+13mn} \\
SPECULOOS            & 1             &                         & \cite{Sebastian+21aa}    \\
ZTF                  &               & 1                       & \cite{Bellm+19pasp}      \\
\midrule 
Total                & 3764          & 51                      &                          \\
\bottomrule
\end{tabular}
\end{specialtable}

The total number of TEPs and transiting BDs discovered by various projects is shown in Table\,\ref{tab:disc}. The number remains dominated by the \textit{Kepler} mission, and by space telescopes in general. The TEPs are categorised according to the source of the data from which the transits were discovered. This is the fairest way to do it whilst avoiding complexity, but does not capture the details of a typical planetary confirmation. The case of WASP-86/KELT-12 provides an illustrative example: it was announced by Faedi et al.\ \cite{Faedi+16} as WASP-86 and, two weeks later, by Stevens et al.\ \cite{Stevens+17aj} under the moniker KELT-12. It is labelled as a WASP planet in TEPCat because precedence was given to the earlier announcement, but the KELT consortium spent a comparable amount of time and effort in their independent discovery and characterisation of the system. The two discovery papers presented very different physical properties, most obviously a difference of a factor of three in the radius of the planet. A subsequent analysis using TESS data \cite{Me22obs1} showed that both discovery papers were inaccurate, primarily because the shallow and long transit was difficult to observe from the ground.

\begin{figure}[H]
\includegraphics[width=13.5cm]{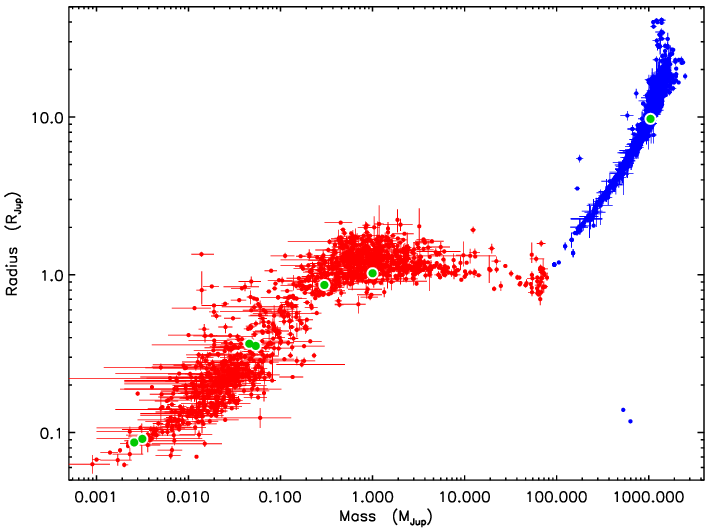} \\
\caption{Mass-radius plot of the objects in TEPCat. Blue points mark the primary objects (stars) and red points the secondary 
objects (TEPs and BDs). In the case of circumbinary planets the primary star in the binary systems is taken to be the primary 
object. The properties of largest solar-system objects are shown in green (without error bars) for illustration.
\label{fig:mr}} \end{figure}

Fig.\,\ref{fig:mr} is a mass--radius plot for both the stellar and planetary components of transiting systems in TEPCat. Both axes are logarithmic due to the large dynamic range of the radii and in particular the masses. The vast majority of points in Fig.\,\ref{fig:mr} follow a clear locus. Whilst the stars show only a small scatter around a main-sequence mass-radius relation, the BDs and planets scatter a lot both above and below the locus: stars are intrinsically more uniform than planets. The scarcity of points at the lower left of the plot arises because such objects are small and low-mass so are difficult to detect and characterise -- the latter effect is the reason why the error bars become large in this area of the diagram. The two primary components with masses of 0.5--0.6\Msun\ but radii around 0.01\Rsun\ are WD 1856+534 \cite{Vanderburg+20natur} and ZTF J2038+2030 \cite{Vanroestel+21apj}, both of which are white dwarfs. The secondary components are a planet of maximum mass 5.2\Mjup\ (WD 1856+534) and a BD of mass $62 \pm 4$\Mjup, respectively; both are on short-period orbits much smaller than the radius of their host star during its post-main-sequence evolution.

\begin{figure}[H]
\includegraphics[width=13.5cm]{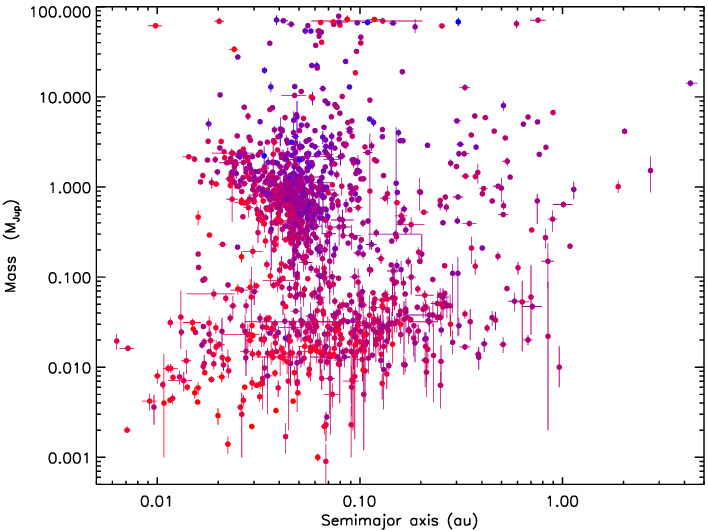} \\
\caption{Masses versus semimajor axes of the relative orbit for the secondary components of systems within TEPCat. 
The points are colour-coded according to host star mass, the most massive (2.38\Msun) being blue and the least massive 
(0.090\Msun) being red. Secondary components of unknown mass or with only an upper limit on their mass are not plotted.
\label{fig:m2a}} \end{figure}

Fig.\,\ref{fig:m2a} shows masses versus semimajor axes for the secondary components of systems within TEPCat, colour-coded on the mass of the host star. Unlike planets found by RV surveys, these are true and not minimum masses. The strongest population is the large and massive planets at short period, i.e.\ the hot Jupiters. The second feature is a population of smaller planets at lower masses, almost all of which were discovered using space telescopes. There is a weaker feature of BDs with masses in the region of 70--75\Mjup\ and separated from the hot-Jupiter population by the BD desert. These objects represent the low-mass end of the star formation process and are classified as BDs on thermonuclear-fusion grounds.

Fig.\,\ref{fig:m2a} is strongly affected by observational and scientific biases. The detection probability decreases to longer orbital periods (larger semimajor axes), because transits are less frequent and the orbital motion of the host star is smaller. Detection probability is also lower for objects less massive than 0.1\Mjup, because they have smaller radii. More unusual objects are over-represented in the plot because they are more likely to be chosen for mass measurement using the limited amount of time available on telescopes with high-precision spectrographs.

\begin{figure}[H]
\includegraphics[width=13.5cm]{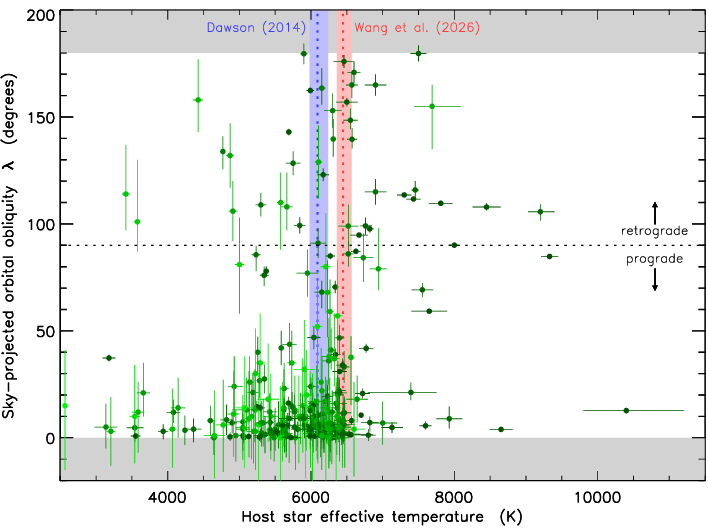} \\
\caption{Sky-projected orbital obliquity ($\lambda$) measurements in TEPCat. The green points are $\lambda$ measurements 
colour-coded according to the sizes of their error bars: more uncertain measurements are shown in lighter green so they do 
not unduly dominate the plot. Only measurements with uncertainties below 30$^\circ$ are plotted. Values of $\lambda$ outside 
$\pm$180$^\circ$ are moved into this interval by adding or subtracting 360$^\circ$. The absolute values of $\lambda$ are 
plotted, and regions outside the interval [0,180$^\circ$] are grey-shaded. The dotted line at $\lambda = 90^\circ$ indicates 
the boundary between prograde and retrograde orbits (labelled). Two proposed boundaries between hot and cool stars are 
plotted for reference: $\Teff = 6090^{+150}_{-110}$ from Dawson \cite{Dawson14apj} and $\Teff = 6447^{+85}_{-119}$ from 
Wang et al.\ \cite{Wang++26apj} (labelled). 
\label{fig:rm}} \end{figure}

In Fig.\,\ref{fig:rm} is plotted the sky-projected orbital obliquity measurements in TEPCat. Orbital obliquity is the angle between the rotational axis of the host star and the orbital axis of the secondary component \cite{Triaud17book,Albrecht++22pasp}. It is usually measured via the Rossiter-McLaughlin effect, which was originally noticed in the eclipsing binary systems $\beta$\,Lyrae \cite{Rossiter24apj} and $\beta$\,Persei \cite{McLaughlin24apj}. It is intrinsically a property of a system, but for convenience the measurements in TEPCat are assigned to individual planets.

Orbital obliquity is governed by the formation of planetary systems, and modified by tidal interactions when planets are in short-period orbits \cite{Triaud11aa,Dawson14apj,Triaud17book,Albrecht++22pasp}. It is therefore an important tracer of star and planet formation and the efficiency of tidal effects. As an example, Albrecht et al.\ \cite{Albrecht+21apj} and Knudstrup et al.\ \cite{Knudstrup+24aa} recently found an excess of systems with nearly polar orbits (i.e.\ $\lambda \sim 90^\circ$). However, subsequent work has not confirmed this result \cite{Siegel++23apj,Knudstrup+24aa}. Several of these papers used TEPCat as their primary source when compiling samples of obliquity measurements.


\section{Classification of transiting systems}
\label{sec:class}

The maintenance of an online database of a certain type of object requires a clear statement about which qualify. In the case of TEPCat these points have been covered in Section\,\ref{sec:scope}. There is, however, a large and expanding literature which uses a wide range of terminology for different kinds of planet (and other object) based on features measurable from observations. In this section we outline a simple classification scheme for various types of planet, attempting to stick to common usage as much as possible.

\begin{figure}[H]
\includegraphics[width=13.5cm]{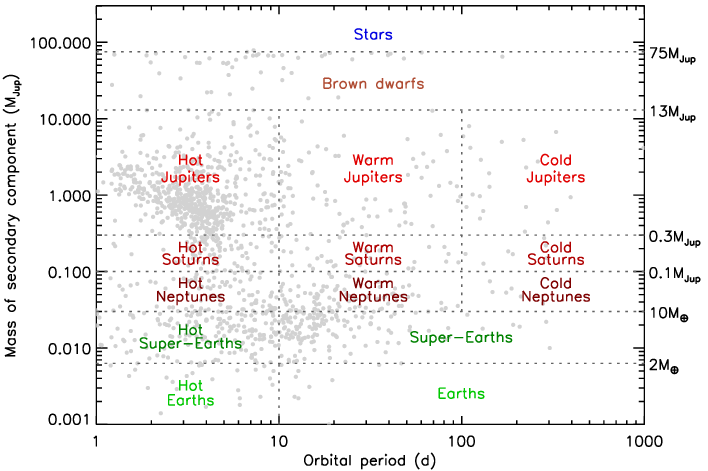} \\
\caption{Schematic diagram for classifying planets according to their masses and orbital periods. 
The various categories of planets are labelled. The boundary masses are shown on the right.
The grey points in the background are the secondary components included in TEPCat.
\label{fig:class}} \end{figure}

Fig.\,\ref{fig:class} shows a simple classification scheme which is based only on the masses and orbital periods of objects, and is closely aligned with nomenclature commonly used in the literature. Stars are more massive than 75\Mjup\ and BDs are 13--75\Mjup\ (see Section\,\ref{sec:scope}). Jupiter-class planets have masses 0.3--13\Mjup, Saturns are 0.1--0.3\Mjup, Neptunes are 10\Mearth\ to 0.1\Mjup, Super-Earths are 2--10\Mearth, and Earth-like planets are less massive than 2\Mearth. Within the planetary regime, planets with an orbital period of less than 10\,d are classified as `hot', those with a period beyond 100\,d as `cold', and those in between as `warm'. The exception to this is that the super-Earths and Earths are subdivided into `hot', but not `warm' and `cold'. This is because such objects are relatively rare and the terminology is not yet widely used; these sub-categories will likely be added in future as and when they become useful. The reason that the temperature divisions have no dependence on mass is that the equilibrium temperature of a planet depends on the host star's temperature and fractional radius, but not on the mass or radius of the planet \cite{Guillot+96apjl,HansenBarman07apj}. One shortcoming of the proposed classification scheme is that it depends on planets' masses, not radii, and many transiting planets (especially the small ones) have measured radii but not masses. This classification scheme is not used in TEPCat (except for the planet/BD/star boundaries) and the interested reader is free to draw up their own scheme.

\begin{figure}[H]
\includegraphics[width=13.5cm]{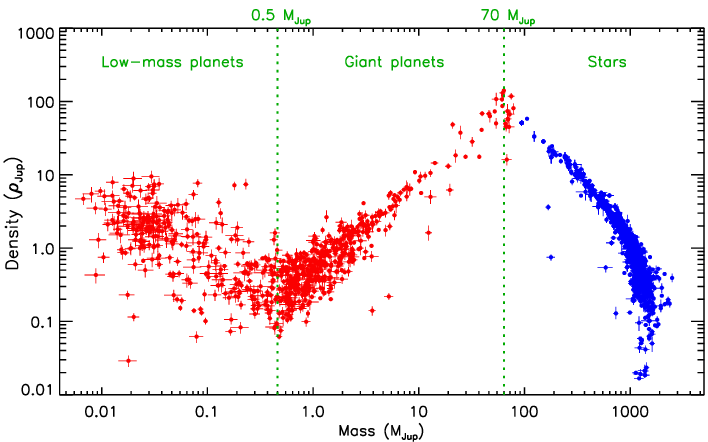} \\
\caption{Mass-density plot inspired by Fig.\,1 of Hatzes \& Rauer \cite{HatzesRauer15apj}. 
The blue points indicate host stars and the red points indicate objects orbiting the host stars.
The three categories have the labels proposed by Hatzes \& Rauer, but revised (larger) boundaries 
are indicated here. Only objects with mass \emph{and} density measurements larger than four times 
their respective error bars are included. The other change versus Hatzes \& Rauer \cite{HatzesRauer15apj} 
is that eclipsing binary systems are not included. \label{fig:mrho}} \end{figure}

A very different approach was taken by Hatzes \& Rauer \cite{HatzesRauer15apj}, who plotted mass versus density for a set of TEPs (taken from TEPCat) and eclipsing binaries \cite{Torres++10aarv}, and divided this into three regimes (low-mass planets, giant planets and stars). It is interesting to check if the trends survive the addition of new data. Fig.\,\ref{fig:mrho} shows a plot of mass versus density for all objects in TEPCat which have measurements of these quantities to a precision of better than 25\% (for both upper and lower error bars), inspired by Fig.\,1 of Hatzes \& Rauer \cite{HatzesRauer15apj}. We followed these authors in dividing the objects into three categories according to how density varies with mass, but revised the boundary masses upwards to 0.5\Mjup\ and 70\Mjup\ (from 0.3\Mjup\ and 60\Mjup), in order to better delineate the categories. The trends of density with mass remain strong with the addition of a further ten years of discoveries, and the pattern for low-mass planets is much clearer. This classification scheme is primarily evolutionary: there is only a small effect due to formation mechanism via chemical composition. It also does not include a distinct BD category, and has not been widely adopted in the literature.


\section{Nomenclature of eclipses}
\label{sec:nomen}

\begin{figure}[H]
\includegraphics[width=13.5cm]{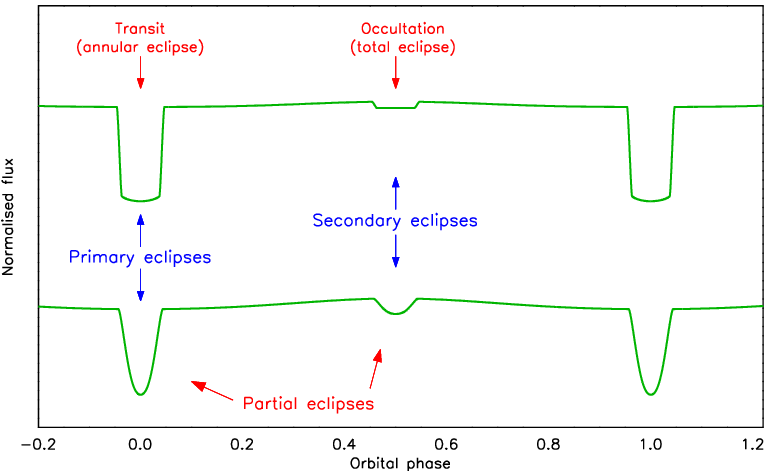} \\
\caption{Schematic of two types of light curve for transiting systems with the correct nomenclature for the eclipses. The light 
curves were created with the {\sc jktebop} code \cite{Me++04mn2,Me13aa} using representative values for the input parameters.
\label{fig:LC}} \end{figure}

Fig.\,\ref{fig:LC} shows two generic light curves of a planetary system. The upper one shows transits and occultations of a typical shape. The lower one shows only partial eclipses due to having a lower orbital inclination (or, equivalently, a larger impact parameter). The \textit{primary eclipses} occur at phase 0.0, and are by definition deeper than the \textit{secondary eclipses} (which is at phase 0.5 if the orbit is circular). The feature in the upper light curve at phase 0.0 is caused by the passage of the dark planet in front of the bright star: it is called a \textit{transit} (from the Latin \textit{transitus}, meaning to `go across' \cite{Oxford10book}) or an \textit{annular eclipse} (from the Latin \textit{annularis}). The feature in the upper light curve at phase 0.5 happens when the planet passes behind the star: it is called an \textit{occultation} (from \textit{occulare} in Latin, meaning `covered over') or \textit{total eclipse} (for obvious reasons). 

It is only possible to have one transit in a system, so the term `primary transit' is a tautology and should be avoided. The alternative phrases `anti-transit' and `secondary transit' are incorrect and should never be used. Similarly, using `eclipse' to mean `secondary eclipse' is ambiguous and should not be done. 

The lower light curve in Fig.\,\ref{fig:LC} is for a system when one object is never totally in front of or behind the other as seen by the observer. In these cases the eclipses are not transits or occultations, but are instead \textit{partial eclipses}. Partial eclipses are relatively rare in planetary systems because planets are almost always much smaller than their host stars, but a small number are known (e.g.\ WASP-67 \cite{Hellier+12mn,Mancini+14aa2}, HIP\,65 \cite{Nielsen+20aa} and WASP-174 \cite{Temple+18mn,Mancini+20aa}). The secondary components are technically \textit{eclipsing} planets, not \textit{transiting} planets, but this distinction is ignored in TEPCat.


\section{Summary and conclusions}
\label{sec:conc}

The search for extrasolar planets has a long history \cite{Struve52obs,BoruckiSummers84icar,CampbellWalker79pasp}. Early claims did not survive further scrutiny \cite{See96aj,Strand43pasp,Vandekamp63aj,Vandekamp69aj,Ribas+18nat} and technological development was needed to enable reliable detections of planets orbiting other stars \cite{MayorQueloz95nat,MarcyButler96apj}. The discovery of the first transiting planet (HD\,209458) occurred in 1999, and a further three years were needed before the second was found (OGLE-TR-56). The discovery rate has since increased hugely, and thousands are now known. 

Scientific studies of a class of objects, such as transiting planets, greatly benefit from the availability of critically-compiled databases of their identities and physical properties. These are needed to keep track of knowledge of these objects, provide samples for parametric studies aiming to explain the physics behind their formation and evolution, give context to new discoveries, and to help develop observing projects and grant applications. TEPCat is a catalogue of the physical and observable properties of transiting planetary systems intended to support all of the possible uses above. It is freely available as both webpages and machine-readable tables, and includes monthly snapshots of the entire database for scientific reproducibility.

Other catalogues exist, including the Extrasolar Planets Encyclopaedia \cite{Schneider+11aa}, the NASA Exoplanet Archive \cite{Christiansen+25psj}, and The Open Exoplanet Catalogue \cite{Rein12xxx}. A detailed overview has been given by Jessie Christiansen \cite{Christiansen18haex}. Whilst there is competition between these catalogues, and with TEPCat, the scientific community benefits from having a choice of which to use. All have their strengths and weaknesses, and direct comparisons can be made to verify their reliability \cite{Bashi++18geo,ChristodoulouDemosthenes20rnaas,Alei+25ac}. In the case of TEPCat, comments and corrections are welcomed by the author because they help to maintain the quality and reliability of the contents.

\funding{JS acknowledges support from STFC under grant number ST/Y002563/1.}

\dataavailability{All data used in this work are freely available from TEPCat at \texttt{https://www.astro.keele.ac.uk/jkt/tepcat/}.}

\acknowledgments{We are grateful to Pierre Maxted, Luigi Mancini and Ahmet Cem Kutluay for comments on a draft of this work. We also wish to thank everyone who has used and cited TEPCat in the past, or has sent comments or corrections for the database.}

\conflictsofinterest{The author declares no conflict of interest.}

\end{paracol}


\reftitle{References}
\externalbibliography{yes}

\end{document}